\documentclass[floatfix,twocolumn,nofootinbib,superscriptaddress,a4paper,preprintnumbers]{revtex4-1}

\usepackage{amsmath}
\usepackage{amssymb}
\usepackage{graphicx}
\usepackage{ulem}
\usepackage[mathscr]{eucal}
\usepackage{color}
\usepackage[T1]{fontenc} 
\usepackage{slashed}
\usepackage{xspace}
\usepackage{hyperref}
\usepackage{xspace}
\usepackage{listings}
\usepackage{gensymb}

\newcommand\SARAH{{\tt SARAH}\xspace}
\newcommand\SPheno{{\tt SPheno}\xspace}

\begin{document}

\preprint{KA-TP-11-2018}

\title{Improved unitarity constraints in Two-Higgs-Doublet-Models}

\author{Mark~D.~Goodsell} 
\email{goodsell@lpthe.jussieu.fr}
\affiliation{Laboratoire de Physique Th\'eorique et Hautes Energies (LPTHE), UMR 7589,
Sorbonne Universit\'e et CNRS, 4 place Jussieu, 75252 Paris Cedex 05, France }
\author{Florian Staub}
\email{florian.staub@kit.edu}
\affiliation{Institute for Theoretical Physics (ITP), Karlsruhe Institute of Technology, Engesserstra{\ss}e 7, D-76128 Karlsruhe, Germany}
\affiliation{Institute for Nuclear Physics (IKP), Karlsruhe Institute of Technology, Hermann-von-Helmholtz-Platz 1, D-76344 Eggenstein-Leopoldshafen, Germany}

\begin{abstract}
Two-Higgs-Doublet-Models (THDMs) are among the simplest extensions of the standard model and are intensively studied in the literature. Using on-shell parameters such as the masses of the additional scalars as input, corresponds often to large quartic couplings in the underlying Lagrangian. Therefore, it is important to check if these couplings are for instance in agreement with perturbative unitarity. The common approach for doing this check is to consider the two-particle scattering matrix of scalars in the large centre-of-mass
energy limit where only point interactions contribute. We show that this is not always a valid approximation: the full calculation including 
all tree-level contributions at finite energy can lead to much more stringent constraints. We show how the allowed regions in the parameter space are affected. 
In particular, the light Higgs window with a second Higgs below $125$~GeV completely closes for large values of the $Z_2$ breaking parameter $|M_{12}|$. We also compare against the loop corrected constraints, which use also the large $\sqrt{s}$ approximation, and find that (effective) cubic couplings are often more important than radiative corrections.   
\end{abstract}
\maketitle

\section{Introduction}
The discovery of a scalar boson  at the Large Hadron Collider with a mass of around 125~GeV was a milestone for particle physics \cite{Aad:2012tfa,Chatrchyan:2012xdj}. This state has all expected properties of the long searched-for Higgs boson, and all particles predicted by the standard model of particle physics (SM) have finally been found. Even if no additional, fundamental scalar has been observed so far at the LHC, it is much too early to give up the possibility that more Higgs-bosons exist which are involved in electroweak symmetry breaking (EWSB). There are several possibilities what the origin and the properties of such states could be. A very attractive and well studied scenario is that a second Higgs doublet exists. After EWSB, the two Higgs doublets yield one particle which has all the properties of the discovered state, but they also predict the presence of one charged and two neutral additional bosons. There exist several constraints on this kind of models: the LHC measurements must be reproduced, the absence of any other signal must be explained, including modifications to rare decay processes. From the theoretical point of view, these models are usually confronted with two conditions: (i) the electroweak vacuum must be stable or at least sufficiently long-lived\cite{Klimenko:1984qx,Velhinho:1994np,Ferreira:2004yd,Barroso:2005sm,Maniatis:2006fs,Ivanov:2006yq,Ivanov:2007de,Ivanov:2008er,Ginzburg:2009dp,Staub:2017ktc}, (ii) unitarity should not be violated\cite{Casalbuoni:1986hy,Casalbuoni:1987eg,Maalampi:1991fb,Kanemura:1993hm,Ginzburg:2003fe,Akeroyd:2000wc,Horejsi:2005da}. In order to probe unitarity in BSM models, the standard procedure in the literature is to calculate the scattering matrix for $2\to 2$ processes involving scalars. Usually, only point interactions are included, which do not vanish for very large scattering energies $\sqrt{s}$. For extensions of the Standard Model, the contributions from scalar trilinear couplings have only been considered for singlet extensions and the minimal supersymmetric standard model\cite{Cynolter:2004cq,Kang:2013zba,Schuessler:2007av}. Therefore, it is time to check if the large $\sqrt{s}$ approximation in THDMs is valid or under which circumstances it might give misleading results. \\
This letter is organised as follows: we show our conventions for THDMs in sec.~\ref{sec:model}, before we briefly summarise our approach to calculate the tree-level unitarity constraints in sec.~\ref{sec:unitarity}. The impact on the parameter space is discussed in sec.~\ref{sec:results}. In sec.~\ref{sec:loop}, we compare against previously derived one-loop results; and rederive the constraints for different unitarity conditions.  We conclude in sec.~\ref{sec:conclusion}.

\section{Model}
\label{sec:model}
The scalar potential of a CP conserving THDM  with softly broken $Z_2$ symmetry reads
\begin{align}
V_{\rm Tree} = & \lambda_1 |H_1|^4 + \lambda_2 |H_2|^4 + \lambda_3 |H_1|^2 |H_2|^2 + \lambda_4 |H^\dagger_2 H_1|^2 \nonumber \\ 
& \label{eq:pot} \hspace{-1cm} + m_1^2 |H_1|^2 + m_2^2 |H_2|^2   +  \left(m_{12} H_1^\dagger H_2 + \frac12 \lambda_5 (H_2^\dagger H_1)^2 + \text{h.c.}\right)
\end{align}
After EWSB, the neutral components of the two Higgs states receive vacuum expectation values (VEVs) of
\begin{equation}
H_i = \left(\begin{array}{c} H_i^+ \\ \frac{1}{\sqrt{2}}\left(\phi_i + i \sigma_i + v_i \right) \end{array} \right) \quad i=1,2
\end{equation}
with $\sqrt{v_1^2+v_2^2} = v\simeq 246$~GeV and $\tan\beta=\frac{v_2}{v_1}$. The mass spectrum consists of superposition of these gauge eigenstates, i.e.
$(\phi_1, \phi_2) \to (h,H)$, $(\sigma_1, \sigma_2) \to (G,A)$ and $(H^+_1, H^+_2) \to (G^+,H^+)$. 
Here, $G$ and $G^+$ are the Goldstone modes of the $Z$ and $W$ boson. The mixing in these sectors is fixed by $\tan\beta$, while in the CP-even sector a rotation angle $\alpha$ defines the transition from gauge to mass eigenstates. In practical applications, one can trade the physical masses $m_h$, $m_H$, $m_A$ and $m_{H^+}$ as well as $\tan\beta$ and $\tan\alpha$ for the quartic couplings. The necessary relations are
{\allowdisplaybreaks
\begin{align}
\label{eq:l1}
\lambda_1 = & \frac{1 + t_\beta^2}{2 (1 + t_\alpha^2) v^2} \left(m_H^2 + m_{12} t_\beta + t_\alpha^2 (m_h^2 + m_{12} t_\beta) \right) \\
 \lambda_2  = & \frac{1 + t_\beta^2}{2 (1 + t_\alpha^2) t_\beta^3 v^2} \left(m_{12} + m_{12} t_\alpha^2 + t_\beta (m_h^2  + m_H^2 t_\alpha^2 ) \right)  \\
 \lambda_3  = & \frac{1}{(1 + t_\alpha^2) t_\beta v^2} \Big[m_h^2 t_\alpha + 2 m_{H^+}^2 (1 + t_\alpha^2) t_\beta \nonumber \\
  & + m_h^2 t_\alpha t_\beta^2 - m_H^2 t_\alpha (1 + t_\beta^2) + m_{12} (1 + t_\alpha^2) (1 + t_\beta^2)\Big] \\
 \lambda_4  = &\frac{1}{t_\beta v^2}\left(-m_{12} + m_A^2 t_\beta - 2 m_{H^+}^2 t_\beta - m_{12} t_\beta^2 \right) \\
 \label{eq:l5}
\lambda_5  = & \frac{1}{t_\beta v^2}\left(-m_{12} - m_A^2 t_\beta - m_{12} t_\beta^2 \right)
\end{align}}
with $t_\beta = \tan\beta$ and $t_\alpha = \tan\alpha$. This has the advantage that physical observables instead of Lagrangian parameters can be chosen as input. However, one needs
to be careful since a randomly chosen set of masses could easily correspond to a problematic set of quartic couplings: for very large couplings perturbativity will be spoilt and also unitarity can be violated.

\section{Unitarity constraints}
\label{sec:unitarity}
Perturbative unitarity constraints come from applying the unitarity of the S-matrix for 
 $2 \to 2$ scalar field scattering amplitudes. We calculate a matrix $a^{ba}_0$ given by
\begin{align}
a_0^{ba} \equiv& \frac{1}{32\pi} \sqrt{\frac{4 |\mathbf{p}^b| |\mathbf{p}^a|}{2^{\delta_{12}} 2^{\delta_{34}}\, s}} \int_{-1}^1 d(\cos \theta) \mathcal{M}_{ba} (\cos \theta),
\label{EQ:a0}\end{align}
which is 
derived proporional to the zeroth partial wave of scattering pairs of scalars  $a$  to pairs $b$ having matrix element $\mathcal{M}(\cos \theta)$, where $\theta$ is the angle between the incoming and outgoing three-momenta ($\mathbf{p}^a, \mathbf{p}^b$ respectively) in the centre-of-mass frame.  
The factor  $\delta_{12} (\delta_{34})$ is $1$ if particles $\{1,2\} (\{3,4\})$ are identical, and zero otherwise. We then find the eigenvalues of this matrix, which we denote $a_0^i$, and insist that they must satisfy
\begin{align}
  |\mathrm{ Re}(a_0^i)| \leq \frac{1}{2}.
\label{EQ:ReLim}\end{align}

\begin{figure}[t]
\includegraphics[width=\linewidth]{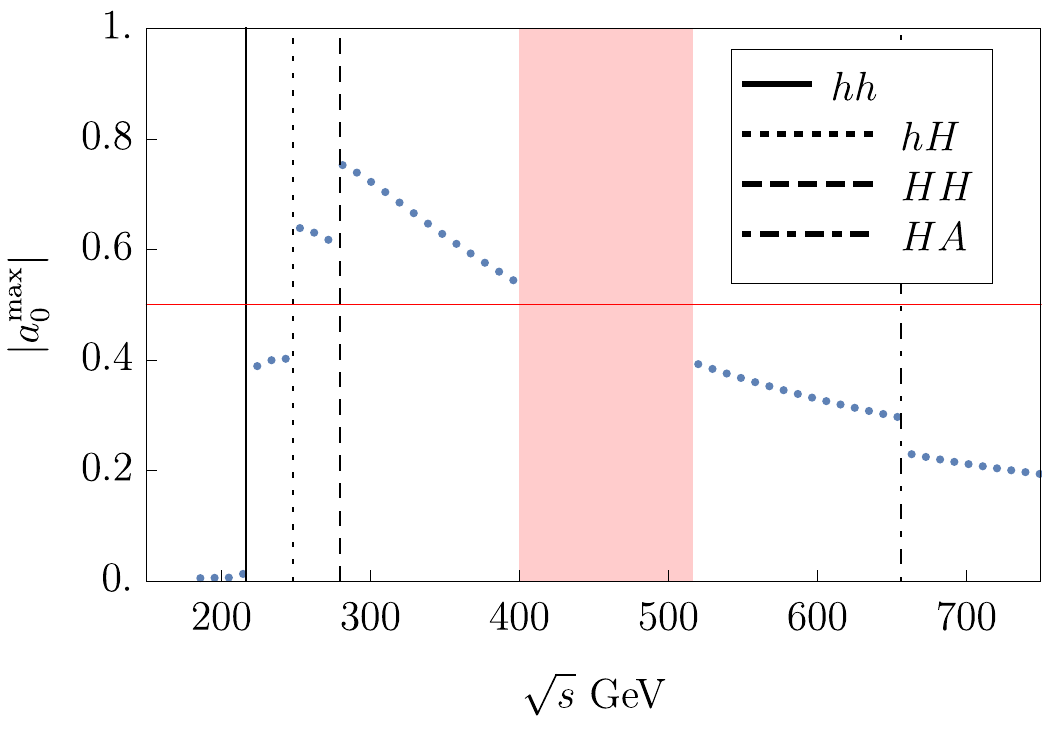} 
\caption{$s$-dependence of the maximal scattering eigenvalue. The black lines indicate the kinematic thresholds while the red region is cut about because of $s$-channel resonance with heavy charged and pseudo-scalar Higgs.}
\label{fig:s}
\end{figure}

Classic unitarity constraints for the THDM have been calculated in the limit of large scattering energies, in which case only the quartic couplings contribute to scattering and the momentum dependence of the prefactor of the integrand in (\ref{EQ:a0}) disappears; moreover all diagrams with propagators are suppressed by the collision energy squared and can be neglected, so the final result appears superficially independent of the scattering energy. This has been applied at tree \cite{} and one-loop \cite{} level. The limits on the quartic couplings at tree level in this approximation are
\begin{align}
\text{Max}\Big\{\left|\lambda_3\pm\lambda_4\right|, \left|\lambda_1+\lambda_2\pm\sqrt{(\lambda_1-\lambda_2)^2+\lambda_4^2}\right|,\left|   \lambda_3\pm\lambda_5\right|,\nonumber\\
\left|3(\lambda_1+\lambda_2)\pm\sqrt{9   (\lambda_1-\lambda_2)^2+(2   \lambda_3+\lambda_4)^2}\right|,\nonumber \\
\left| \lambda_3+2   \lambda_4\pm3 \lambda_5\right|,
\left|   \lambda_1+\lambda_2\pm\sqrt{(\lambda_1-\lambda_2)^2+\lambda_5^2}\right|\Big\}< 8 \pi. 
\label{EQ:Classic}\end{align}
However, it has not been tested if the large $s$ approximation is valid in all BSM models in which it is applied. It could be that large contributions are present at smaller $s$ which then rule out given parameter regions in the considered model. The theory could develop a Landau pole before $s$ is sufficiently large to neglect the masses, or could be defined with a low cutoff. And at large values of the couplings, their running is usually sufficiently fast so that the values of the couplings at an energy scale $\sqrt{s}$ are vastly different from those at lower energies. So in order to be able to test unitarity at finite $s$, the {\tt Mathematica} package \SARAH has now been extended. The salient features are: (i) all tree-level diagrams with internal and external scalars are included to calculate the full scattering matrix; (ii)  We neglect all gauge couplings, and treat Goldstone bosons as physical particles with mass equal to the gauge boson; (iii) the calculation is done in terms of mass eigenstates, i.e. the full VEV-dependence is kept; (iv) the numerical evaluation is done with the Fortran code \SPheno \cite{Porod:2003um,Porod:2011nf}; (v) large enhancements close to poles are cut in order not to overestimate the limits. This is demonstrated at one-example in Fig.~\ref{fig:s}. More details and derivations of our full procedure are given in the accompanying paper \cite{inprepsarah}.

\section{Results}
\label{sec:results}
\begin{figure*}[htb]
 \includegraphics[width=0.32\linewidth]{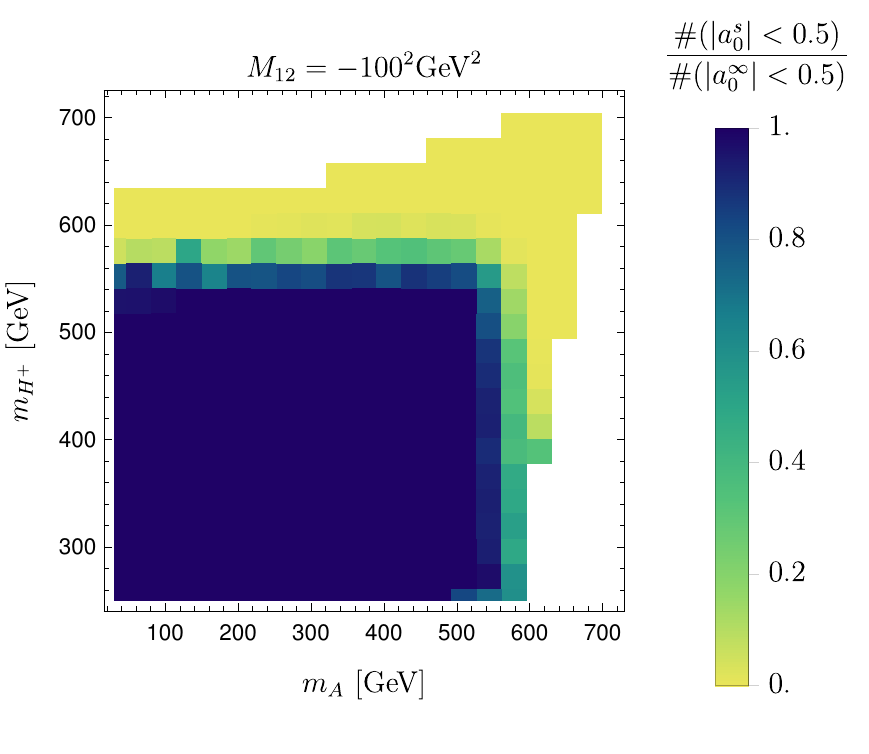} \hfill 
 \includegraphics[width=0.32\linewidth]{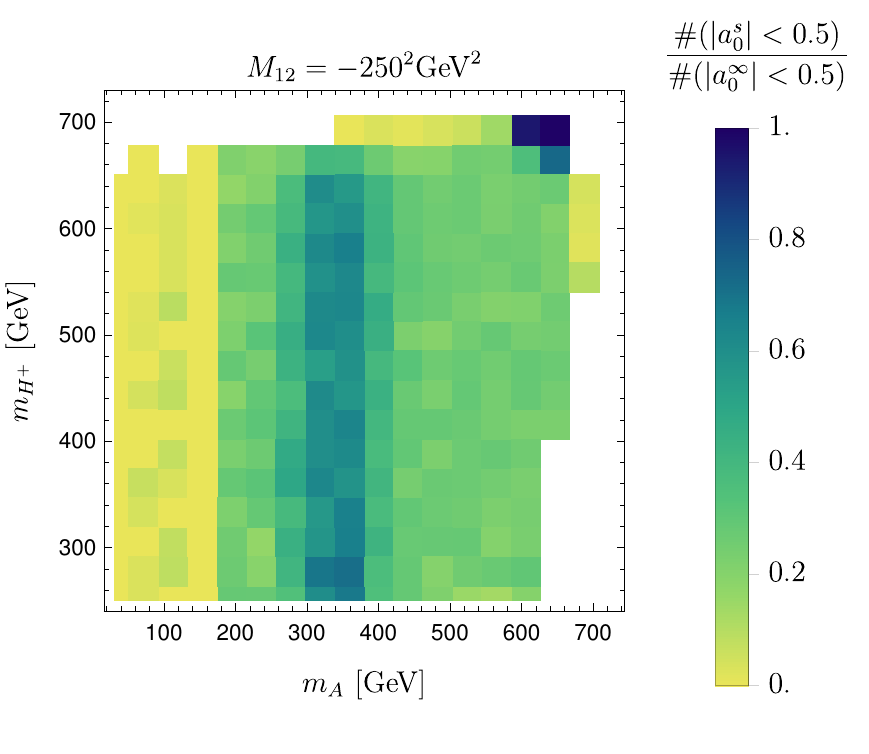} \hfill
 \includegraphics[width=0.32\linewidth]{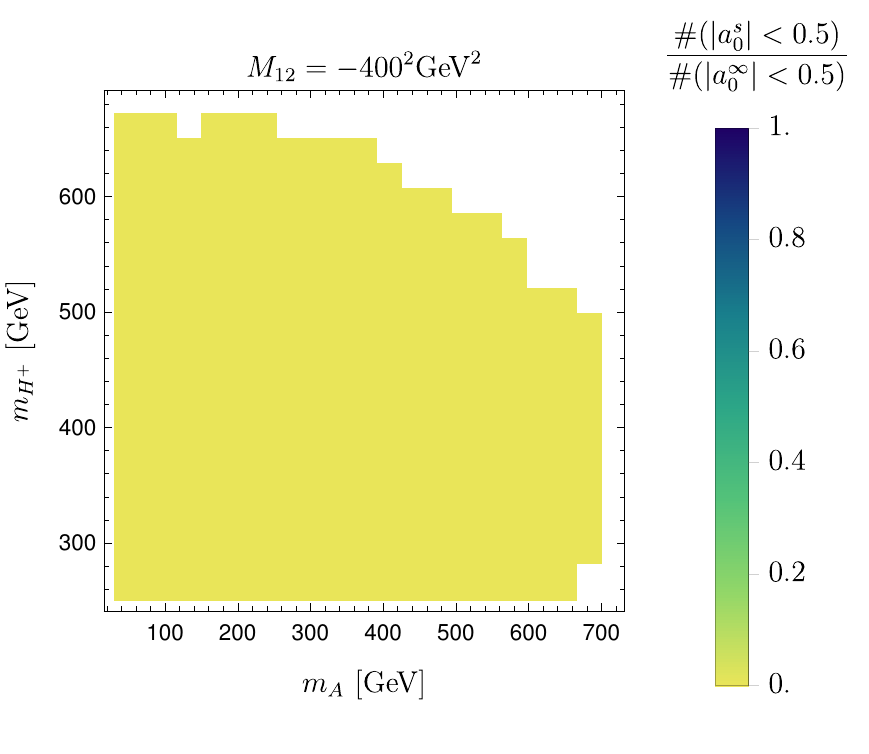} \\
 \includegraphics[width=0.32\linewidth]{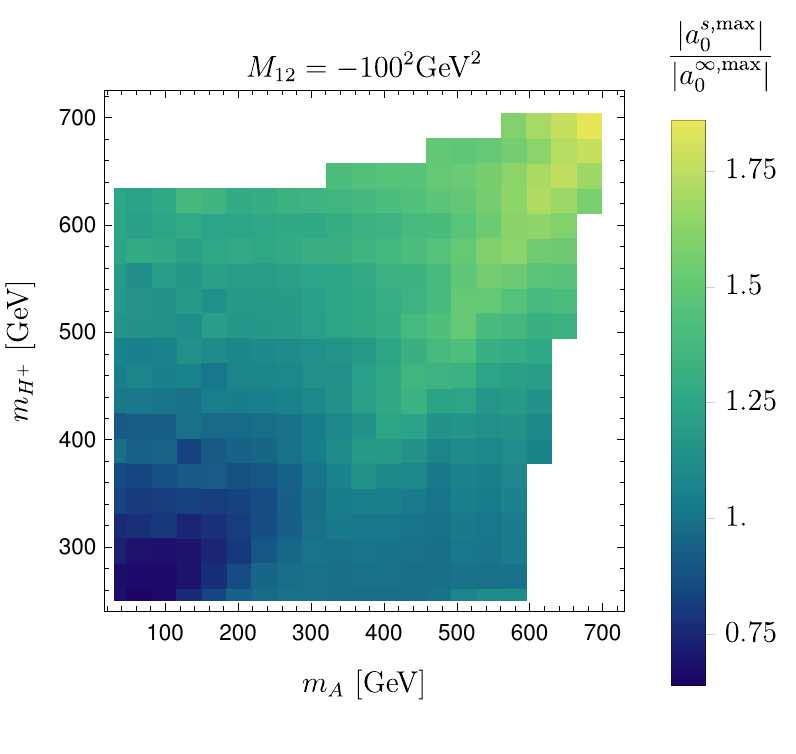} \hfill 
 \includegraphics[width=0.32\linewidth]{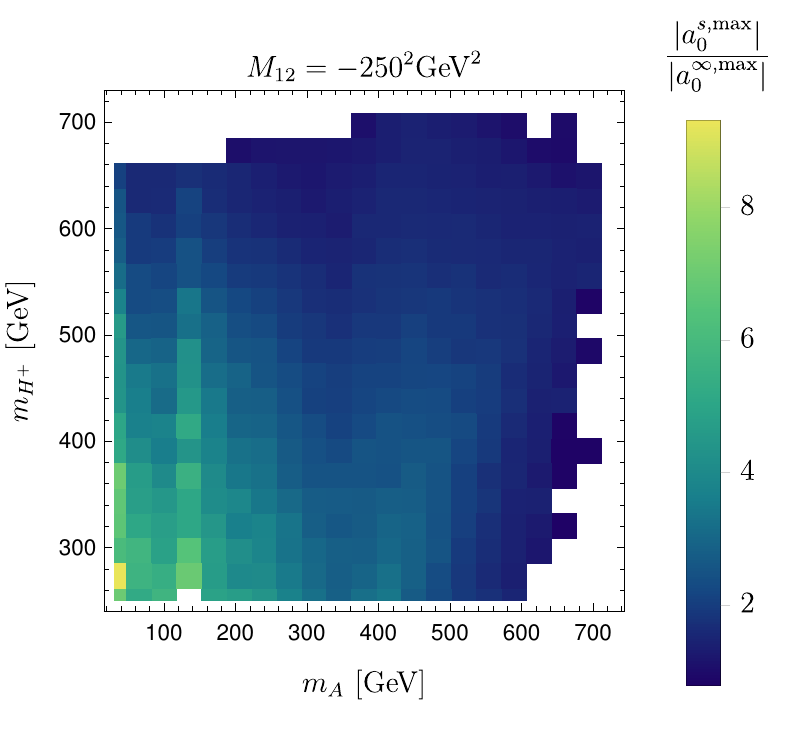} \hfill
 \includegraphics[width=0.32\linewidth]{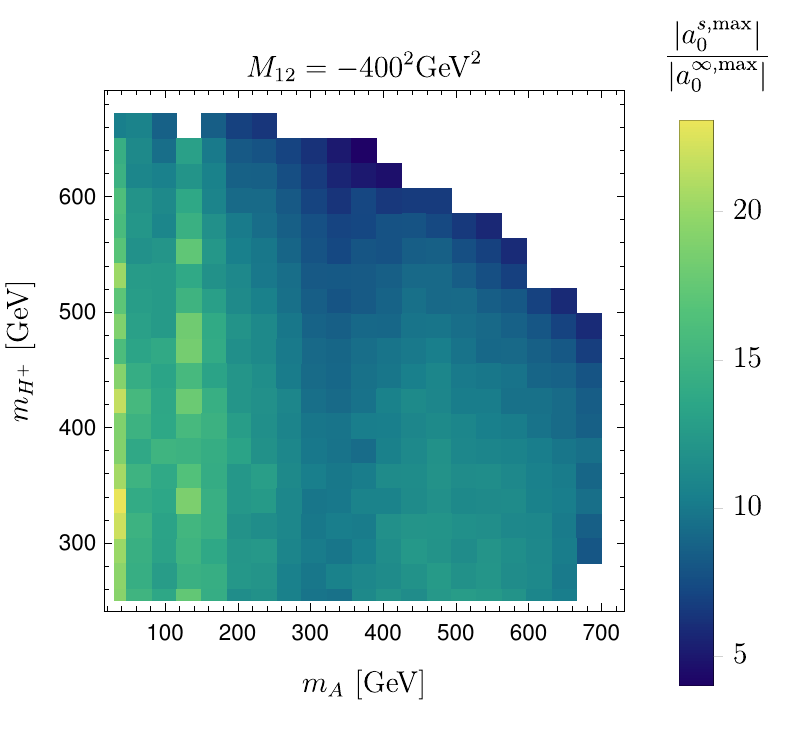} 
\caption{Comparison between the old and new unitarity constraints for a second light CP even scalar for three different values of $M_{12}$. The figures in the first row show the ratio of points 
which pass the old unitarity constraints but are ruled out by the new ones. The second row shows the average enhancement in the maximal scattering element.
The other parameters were varied in the ranges $m_H \in [60,120]$~GeV, $m_A \in [30,1000]$~GeV, $m_{H^+} \in [250,1000]$~GeV,  $\tan\alpha \in [-0.25,-1.5]$, $\tan\beta \in [1,2]$.}
\label{fig:light}
\end{figure*}
In this section we shall study the impact of the improved unitarity constraints on the two Higgs doublet model at tree level. We have chosen for our discussion type--I, but the results hold also for 
other models, becuase our we omit fermions from our scattering processes. Hence there is only an indirect difference between the constraints for type--I and type--II: the limits from flavour observables are stronger for light charged Higgs masses for type--II. Hence, the $m_{H^+}$ must be larger in general for type--II \cite{Misiak:2017bgg}. On the other hand, we include the  constraints from Higgs searches via {\tt HiggsBounds} \cite{Bechtle:2008jh,Bechtle:2011sb,Bechtle:2013wla}, which can vary to a lesser extent between type I and II models.

Our numerical analysis is based on the \SPheno~\cite{Porod:2003um,Porod:2011nf} interface of \SARAH~\cite{Staub:2008uz,Staub:2009bi,Staub:2010jh,Staub:2012pb,Staub:2013tta}.  
By default, {\tt SPheno} calculates the mass spectrum  at the full one-loop level  and includes all important two-loop corrections to the neutral scalar masses \cite{Goodsell:2014bna,Goodsell:2015ira,Braathen:2017izn}. However, we shall not make use of these routines in the following but work at tree-level, or equivalently under the assumption that an OS calculation works in principle (with all the caveats discussed in Ref.~\cite{Krauss:2017xpj}). This is because we cannot (yet) calculate quantum corrections to unitarity at finite $s$, and when the couplings are large in almost all cases the quantum corrections to masses/couplings become very large: this gives further motivation for including only constraints at finite $s$!

\subsection{The light Higgs window}
We start with a discussion of the effects in the case that both CP even Higgs states have masses of 125~GeV or below. A comparison 
between the `classic' -- equation (\ref{EQ:Classic}) --  and new constraints is given in Fig.~\ref{fig:light}. Obviously, one finds much stronger constraints in two different cases once finite $s$ is considered: (i) for smallish $|M_{12}|$ the wedge $M_A=M_{H^+} \gg m_H$ disappears; (ii) for larger $|M_{12}|$ the scattering amplitude in the overall $(m_A,m_{H^+}$ grows signifcantly. The responsible channels 
and best scattering energies causing these effects are quite different:
\begin{itemize}
 \item Small $|M_{12}|$: consider the following simplified hierarchy, 
 \begin{equation}
 m_H = m_h = -\sqrt{|M_{12}|} \ll m_A = m_{H^+} \sim \sqrt{s}
\end{equation}
together with $\tan\beta=-1/(\tan\alpha)=1$. 
The dominant channels are those with heavy external states and a light Higgs exchange. For instance, the amplitude $AA\to AA$ can be approximated as
\begin{equation}
a_0(AA\to AA) = \frac{m_A^4 \left(-2 s \log \left(\frac{m_h^2}{-4 m_A^2+m_h^2+s}\right)\right)}{8 \pi  s v^2 \sqrt{s \left(s-4 m_A^2\right)}} 
\end{equation}
From that, we get that the ratio compared to the old constraints 
\begin{equation}
\frac{a_0^{\rm max}}{a_0^{s\to\infty}} =-\frac{4 m_A^2 \log \left(\frac{m_h^2}{-4 m_A^2+m_h^2+s}\right)}{3 \sqrt{s \left(s-4 m_A^2\right)}}
\end{equation}
This ratio becomes maximal slightly above the kinematic threshold $s_{\rm Threshold} = 4 m_A^2$ and an enhancement of 2--3 is possible. Thus, the best scattering energy $\sqrt{s}$  is around 1--2~TeV. 
\item Large $|M_{12}|$: in this case we can consider the following, simplified  hierarchy
\begin{equation}
 m_H = m_h \sim \sqrt{s} \ll m_A=m_{H^+}=\sqrt{|M_{12}|}
\end{equation}
Now, the dominant scattering processes are those with light external scalars only. The maximal eigenvalue of the full scattering matrix is roughly given 
by diagonalising the submatrix with CP-even states only
\begin{equation}
\begin{pmatrix}
hh\to hh & hh\to HH & hh\to hH \\
 & HH\to HH & HH \to hH \\
  & & hH\to hH
\end{pmatrix}
\end{equation}
By doing that, we find that the ratio between the old and new results scales as 
\begin{equation}
\frac{a_0^{\rm max}}{a_0^{s\to\infty}} =\frac{2 M_{12} \log \left(\frac{m_h^2}{s-3 m_h^2}\right)}{\sqrt{s \left(s-4 m_h^2\right)}} \sim \frac{1}{2}\frac{|M_{12}|}{m_h^2}
\end{equation}
Thus, this ratio grows very quickly with increasing $M_{12}$ and one finds very strong unitarity constraints already at scattering energies $\sqrt{s}$ of a few hundred GeV. 
\end{itemize}

\subsection{Heavier Scalar}
\paragraph{Stronger Constraints}
We turn now to the case that all new scalars are heavier than the SM-like Higgs. We start with a short analytical estimate for parameter regions in which difference between the our calculation
and previous results show up. 
\begin{figure}[tb]
 \includegraphics[width=\linewidth]{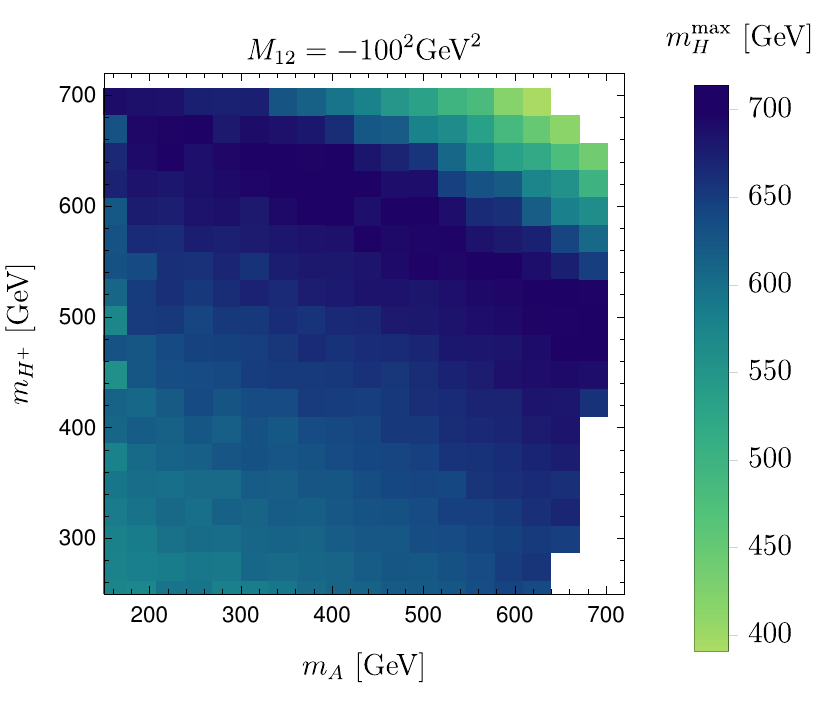} \\
 \includegraphics[width=\linewidth]{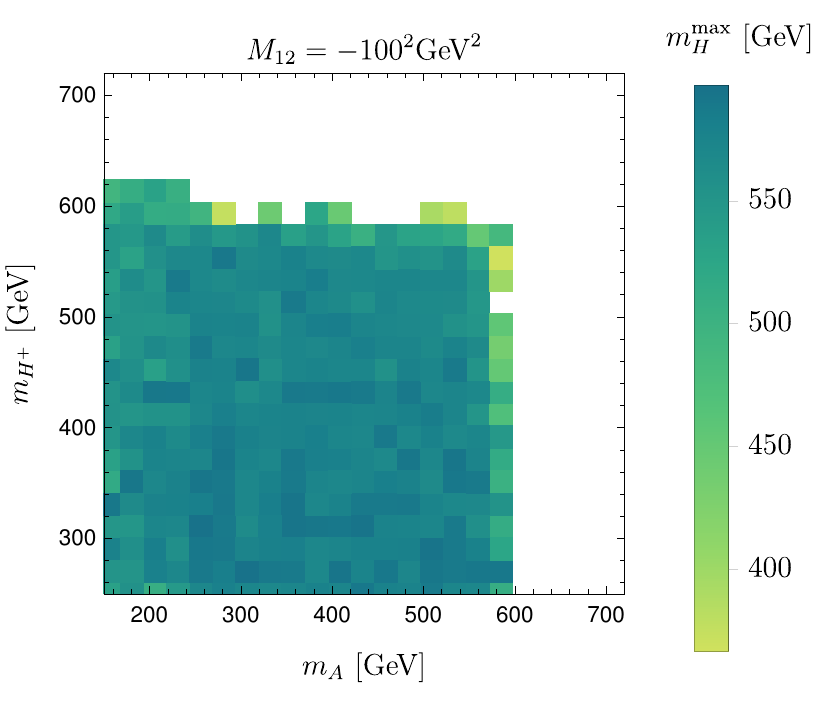}  
\caption{The maximal value of $m_{H^+}$ when using the large $s$ approximation (first row) or the full calculation (second row). Here, we varied $m_{H^+} \in [250,1000]$~GeV, $M_{12} \in [-100^2,-1000^2]~\text{GeV}^2$, $\tan\alpha \in [-0.25,-1.5]$, $\tan\beta \in [1,2]$.}
\label{fig:maxMH}
\end{figure}
This is, for instance, the case for the configuration
\begin{equation}
m_A \sim M_H \sim M_{H^+} \gg \sqrt{|M_{12}|}.
\end{equation}
Assuming again the $\tan\beta=-1/\tan\alpha=1$ for the moment, the maximal eigenvalue for the scattering matrix in the large $s$ limit is
\begin{equation}
a_0^{{\rm max},s\to\infty} = \frac{1}{16\pi v^2} (8 M_{12} + 4 m_A^2 + 5 m_h^2) \simeq \frac{1}{4\pi v^2} m_A^2
\end{equation}
We want to compare this with the scattering $HA \to HA$ scattering process which includes diagrams with the SM-like Higgs in the propagator. We find
\begin{align}
 a_0(HA\to HA) & \simeq  \frac{m_h^2}{16 \pi  v^2 \sqrt{s \left(s-4 m_A^2\right)}} \times \Big[\left(4 m_A^2-s\right) \nonumber \\
& \hspace{-1cm}-\left(4 M_{12}+2 m_A^2+m_h^2\right)^2 \log \left(\frac{m_h^2}{-4 m_A^2+m_h^2+s}\right)\Big]  \\
  & \simeq  \frac{m_A^4 \log \left(\frac{m_h^2}{-4 m_A^2+s+m_h^2}\right)}{4 \pi  v^2 \sqrt{s \left(s-4 m_A^2\right)}}
\end{align}
Thus, for $s=5 m_A^2$ close to the kinematic threshold  we find an enhancement of roughly $|\frac{2}{\sqrt{5}}\log\frac{m_h}{m_A}|$ compared to the large $s$ approximation. For $m_A=700$~GeV this corresponds to nearly a factor of 2. We can confirm this by making use of the full numerical machinery. In Fig.~\ref{fig:maxMH} we show the impact on the maximal allowed value for $m_H$ in the $(m_A,m_{H^+})$ plane while scanning over all other parameters as indicated in the caption.  We see that this value shrinks significantly and a large region of the plane which is allowed by the old constraints is no longer accessible.

\paragraph{Weaker Constraints}
If we consider the scattering up to a finite $\sqrt{s}$, 
we can find that the scatter eigenvalues become smaller compared to the limit $\sqrt{s}\to\infty$ for several reasons: (i)
the dominant channels can be kinematically forbidden; (ii) there can be a negative interference between the point interactions and the propagator diagrams; (iii)
the dominant channels can be cut out because of possible resonances in order not to overestimate the unitarity constraints. 
\begin{figure}[tb]
\includegraphics[width=\linewidth]{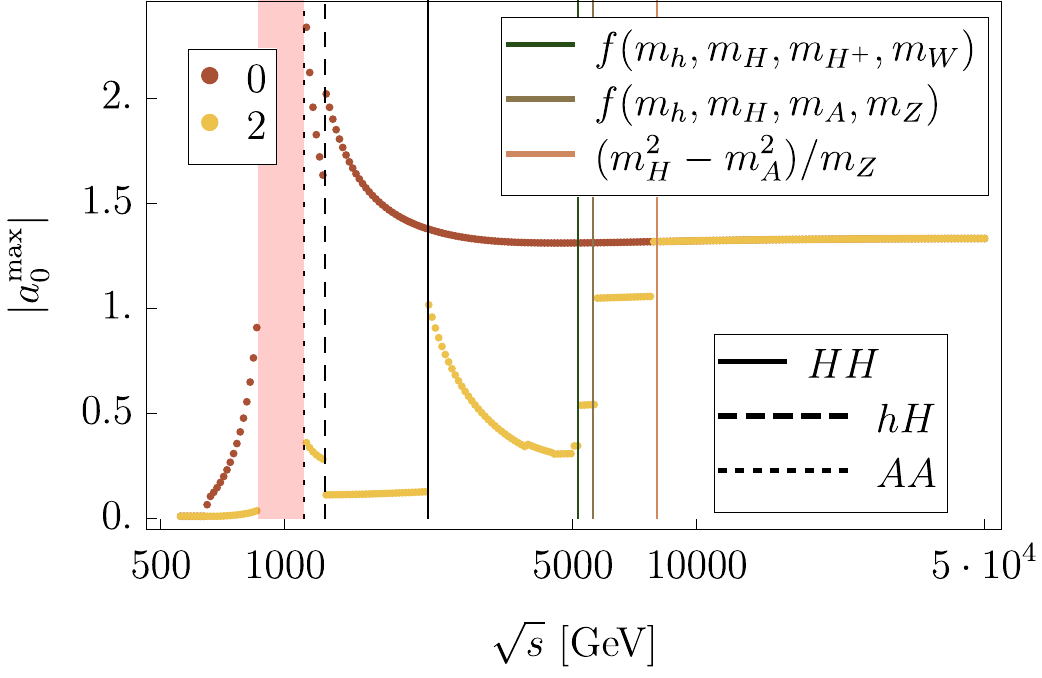} 
\caption{$s$-dependence of the maximal scattering eigenvalue. Here, we have used two possibilities how to deal with the $t$- and $u$-channel poles. The black lines indicate the kinematic thresholds while the red region is cut about because of $s$-channel resonances. The vertical lines indicate limits at which $t$- and $u$-pole disappear. The two options for dealing with a $t$/$u$-channel poles are: (i) the poles are ignored and the full scattering-matrix is taken into account (option 0); (ii) the elements affected by the poles are dropped and a partial diagonalisation of the remaining matrix is performed (option 2). }
\label{fig:tupoles}
\end{figure}
Due to these effects, one needs to ask the question to which energy scale we have actually probed scattering processes of scalars at the LHC. Of course, the LHC is running 
with $\sqrt{s}=14$~TeV. However, it is unrealistic to assume that the full energy is available in the $2\to 2$ scattering of scalars. Moreover, there are different options to handle the $t$- and $u$-channel poles, which can appear if internal states become on-shell, depending on how aggressive or conservative the limits should be: if we remove these poles either completely or only by a partial diagonalisation of the scattering matrix, 
large contributions to the scattering can be dropped at small $s$. We demonstrate via one example in Fig.~\ref{fig:tupoles} where the maximal eigenvalue as a function of $\sqrt{s}$ is shown. If we completely ignore the $t$- and $u$ poles we see a huge enhancement close to some kinematic thresholds. In contrast, if we work with a partial diagonalisation as proposed in Ref.~\cite{Schuessler:2007av} we see that we find  the eigenvalue of the large $s$ approximation only for $\sqrt{s}>10$~TeV. This might be rather surprising since all involved masses are below 1~TeV!

\section{Comparison with loop corrections}
\label{sec:loop}
Since one of our motivations for considering finite $s$ scattering is that the quantum corrections to masses and couplings become large as we increase the scattering energy, it is also important to examine the effect of loop corrections to unitarity. Moreover, the boundary of unitarity may also coincide with a loss of perturbativity. In general, loop corrections to unitarity have been very little studied in BSM models; however, in the context of the THDM, they were considered in Ref.~\cite{Grinstein:2015rtl} in the limit of $\sqrt{s}$ much larger than the masses in the theory. We can therefore make a direct comparison. In that paper, they presented general formulae for the loop corrections to $a_0$ in terms of the quartic couplings of the theory evaluated at the scale $\sqrt{s},$ which are effectively independent of the particle masses.  Results for two scenarios, one with an $SO(3)$ symmetry and another with ``MSSM-like'' couplings, were presented.

We shall make our comparison with the ``MSSM-like couplings''; in {\tt SARAH} conventions this means
\begin{align}
\lambda_1 = \lambda_2, \qquad \lambda_4 = - \lambda_3 - 2 \lambda_1, \qquad \lambda_5 = 0.
\end{align}
With these restrictions the `classic' tree-level constraints of equation (\ref{EQ:Classic}) simplify to
\begin{align}
|8 \lambda_1 - \lambda_3 | \le 8 \pi, \qquad |2\lambda_1 + 2\lambda_3 | \le 8 \pi, 
\end{align}
which describe a rhombus inlcuding the origin. Requiring stability of the potential requires
\begin{align}
\lambda_1 > 0, \qquad \lambda_3 > -2 \lambda_1 ,
\end{align}
which, when we combine the two, leaves a portion of the parameter space where $\lambda_1$ is at most $\frac{4}{3}\pi$, and $\lambda_3 < 4\pi. $ 

\begin{figure}\centering
  \includegraphics[width=0.48\textwidth]{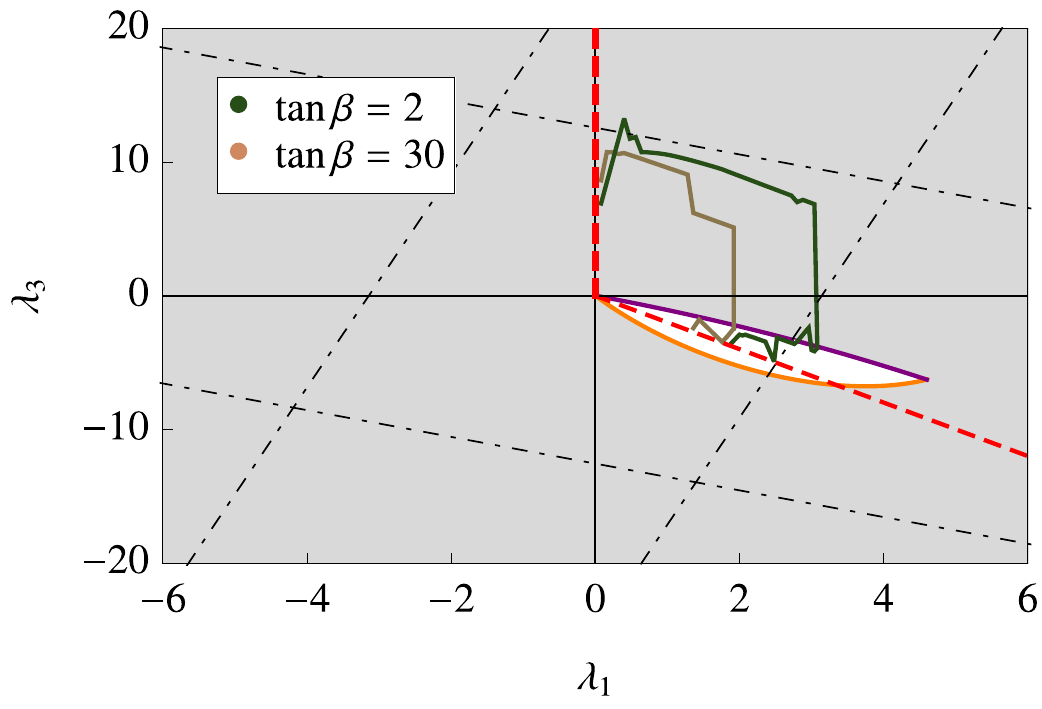} \\
  \includegraphics[width=0.48\textwidth]{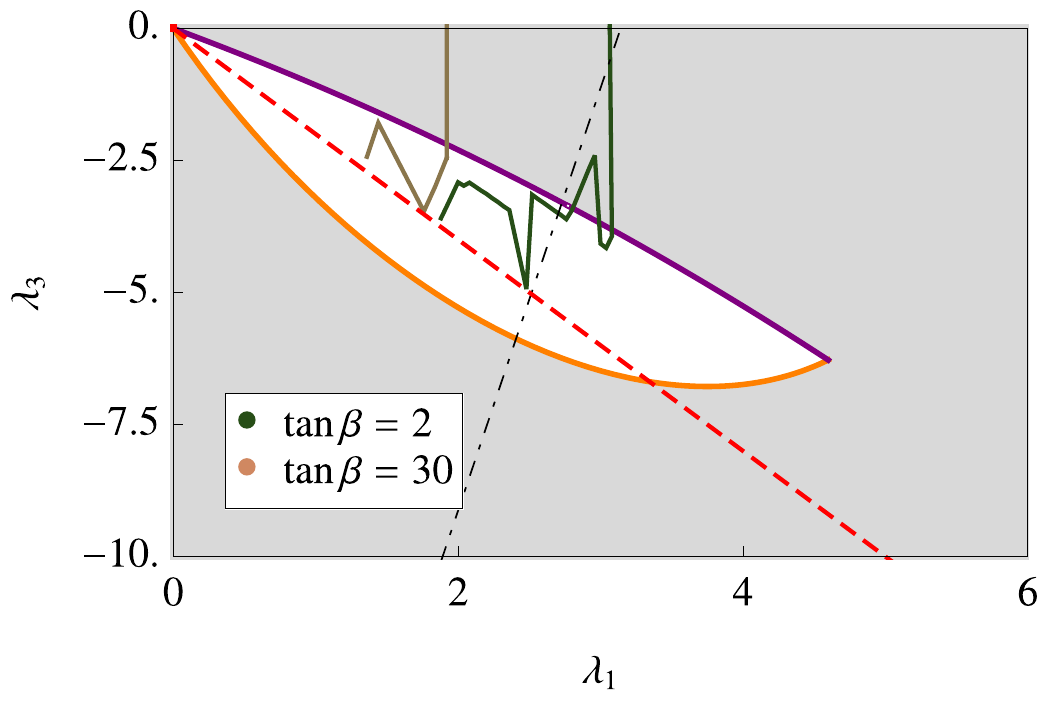}
  \caption{Tree-level and one-loop constraints on $\lambda_1$ and $\lambda_3$ in the ``MSSM-like'' THDM. Quartic-only tree-level constraints are shown as black dot-dashed lines, the vacuum stability constraint $\lambda_3 \ge - 2 \lambda_1$ is the red dashed line; our tree-level contraints including trilinears are labelled with $\tan \beta = 2$ and $\tan \beta =30$. 
  The one-loop allowed region from \cite{Grinstein:2015rtl} is the white region enclosed by the solid purple and orange curves. The second plot is a zoom into the first one. }
\label{FIG:Grinstein}
\end{figure}

In the previous sections, we applied the unitarity constraint $| \mathrm{Re} (a_0^i) | < 1/2$, but in \cite{Grinstein:2015rtl} they apply a different constraint, which we shall now examine. The starting point is the equation
\begin{align}
  \mathrm{Im} (a_0^{i}) \le& |a_0^{i}|^2,
\label{EQ:optical}
\end{align}
(for an elementary derivation see \cite{inprepsarah}). 
Naively this gives simply $|a_0^i| \le 1$, which is a constraint sometimes appllied, but with a little rearranging we have
\begin{align}
\mathrm{Re} (a_0^i)^2 \le |\mathrm{Im} (a_0^i) | ( 1 - |\mathrm{Im} (a_0^i) |)
\end{align}
which gives the classic limit (\ref{EQ:ReLim}). This limit makes no assumption of perturbativity, and indeed when $\mathrm{Re}(a_0^i)$ obtains its maximum value then $\mathrm{Im} (a_0^i) = |\mathrm{Re}(a_0^i)|.$ Since $\mathrm{Im}(a_0^i)$ is only generated at first at one loop order, then saturating this bound would potentially require violating perturbativity. On the other hand, rearranging again, we can write the above as
\begin{align}
| a_0^i - \frac{i}{2}|^2 \le \frac{1}{4}.
\label{EQ:LoopConstraint}\end{align}
If we have complete ignorance of $\mathrm{Im} (a_0^i)$ then we just recover the same constraint as above. However, if we have calculated $a_0$ at one loop and assume that perturbativity holds, then we can use our calculated values for the real and imaginary parts of $a_0^i$ and use the above constraint. Focussing on one eigenvalue, let us write 
\begin{align}
a_0^i \equiv& a_0^0 + b_R + i b_I
\end{align}
and expand eq.~(\ref{EQ:LoopConstraint}) then we find
\begin{align}
(a_0^0)^2 + b_R^2 + 2 a_0^0 b_R - b_I + b_I^2 \le 0.
\end{align}
Now Ref.~\cite{Grinstein:2015rtl} then appeal to perturbation theory so that
\begin{align}
b_I = (a_0^0)^2 + \mathrm{higher\ order\ terms}
\label{EQ:BadAssumption}\end{align}
and then obtain
\begin{align}
2 a_0^0 b_R \le - b_R^2 + ... \rightarrow |a_0^0 | \ge \frac{1}{2} |b_R|.
\label{EQ:GrinConstraint}\end{align}
This can then be a very strong constraint. In Fig.~\ref{FIG:Grinstein} we show the constraints from applying eq.~(\ref{EQ:LoopConstraint}) as done by Ref.~\cite{Grinstein:2015rtl}, with the constraints from our trilinear couplings and the tree-level constraints for comparison. The tree-level quartic-only and one-loop constraints are independent of $\tan \beta$ and all of the mass scales (except that they should be interpreted as couplings evaluated at a renormalisation scale $\sqrt{s}$), whereas for our scan we choose two values of $\tan \beta$ (marked on the plot) and fix the tree-level lightest Higgs mass to be $125$ GeV -- this is enough to determine all of the remaining free parameters once $\lambda_1$ and $\lambda_3$ are specified.  

We see from Fig.~\ref{FIG:Grinstein} that even though the loop-level constraints seem extremely severe, our tree-level trilinear constraint still removes a significant chunk of the remaining parameter space.

However, these one-loop constraints have the curious feature of excluding couplings near the origin, which arises from the regions where one scattering eigenvalue vanishes \emph{at tree level}. Indeed, from eq.~(\ref{EQ:GrinConstraint}) we see that if $a_0^0 =0$ (as can happen for linear combinations of the couplings) then unitarity is apparently violated. In the notation of Ref.~\cite{Grinstein:2015rtl} the purple curve corresponds to the eigenvalue $a_{0}^{110\mathrm{odd}} $, which derives from the scattering of
\begin{align}
\epsilon_{\alpha \beta} \Phi_1^\alpha \tau^3 \Phi_2^\beta \rightarrow  \epsilon_{\alpha \beta} \Phi_1^\alpha \tau^3 \Phi_2^\beta
\end{align}
where $\tau^3 = \frac{1}{2} \left(\begin{array}{cc} 1 & 0 \\ 0 & -1 \end{array}\right) $ and  gives the scattering eigenvalue at tree level of
  $$a_0^0 = 2\lambda_1 + 2\lambda_3.$$
 
The orange curve corresponds to $a_{0-}^{000\mathrm{even}}$  from scattering
\begin{align}
  \Phi_i^\dagger \Phi_i \rightarrow  \Phi_j^\dagger \Phi_j ,  \qquad i =\{1,2 \},
\end{align}
which give the scattering eigenvalues at tree level of
$$a_0^0 = \{- 8 \lambda_1 + \lambda_3, -4 \lambda_1 - \lambda_3 \}. $$

We therefore see that the one-loop constraints arise starting from the lines $\lambda_1 + \lambda_3 = 0 $ and $4 \lambda_1 + \lambda_3 =0. $ The reason for this is, however, assuming that the higher-order terms in eq.~(\ref{EQ:BadAssumption}) are not important. Indeed, in the cases where $a_0^0 = 0$ for $\lambda_i \ne 0$ we would apparently badly violate perturbation theory -- but this is just because we have only computed up to one loop, and have a tuned cancellation at tree level. Since eq.~(\ref{EQ:GrinConstraint}) compares a tree-level and one-loop amplitude this seems particularly bad. Hence, if we examine the perturbation series more closely, specialising to the case of only quartic couplings for simplicity, and define $\lambda$ to be a number of $\mathcal{O} (\lambda_i)$ as a perturbation series parameter, so that 
\begin{align}
  b_R \equiv& b_{1,R} \lambda^2 +  b_{2,R} \lambda^3 + ... \nonumber\\
  b_I \equiv& (a_0^0)^2 \lambda^2 + b_{2,I} \lambda^3 + ...
\end{align}
we see that $a_0^{2\rightarrow n}$ is nonzero first for $2\rightarrow 4$ processes at order $\lambda^2$. Hence defining $\sum_{n>2} | a_{2 \rightarrow n}|^2 \equiv |X|^2 \lambda^4$ we have, order by order in perturbation theory up to $\lambda^4$:
\begin{align}
  2 a_0^0 b_{1,R} - b_{2,I} =& 0 \label{EQ:Better}\\
  (a_0^0)^4 + b_{1,R}^2 + 2 a_0^0 b_{2,R} - b_{3,I} + |X|^2 =& 0.
\end{align}
We see that the origin of eq.~(\ref{EQ:GrinConstraint}) depends on neglecting $b_{2,I}$, but if we include the information from eq.~(\ref{EQ:Better}) then we would have obtained instead of eq.~(\ref{EQ:GrinConstraint}):
\begin{align}
  b_R^2 + \mathrm{higher\ order\ terms\ of\ indeterminate\ sign} =0. \nonumber
\end{align}
Furthermore, when $a_0^0=0$ we simply recover $b_{3,I} \ge 0$ and $b_{2,I} =0$, which we could surmise from $a_0 $ being of $\mathcal{O}(\lambda^2) $ and the standard unitarity relation. We do not obtain any new constraint beyond $|\mathrm{Re}(a_0)| \le \frac{1}{2}.$

\begin{figure}\centering
  \includegraphics[width=0.48\textwidth]{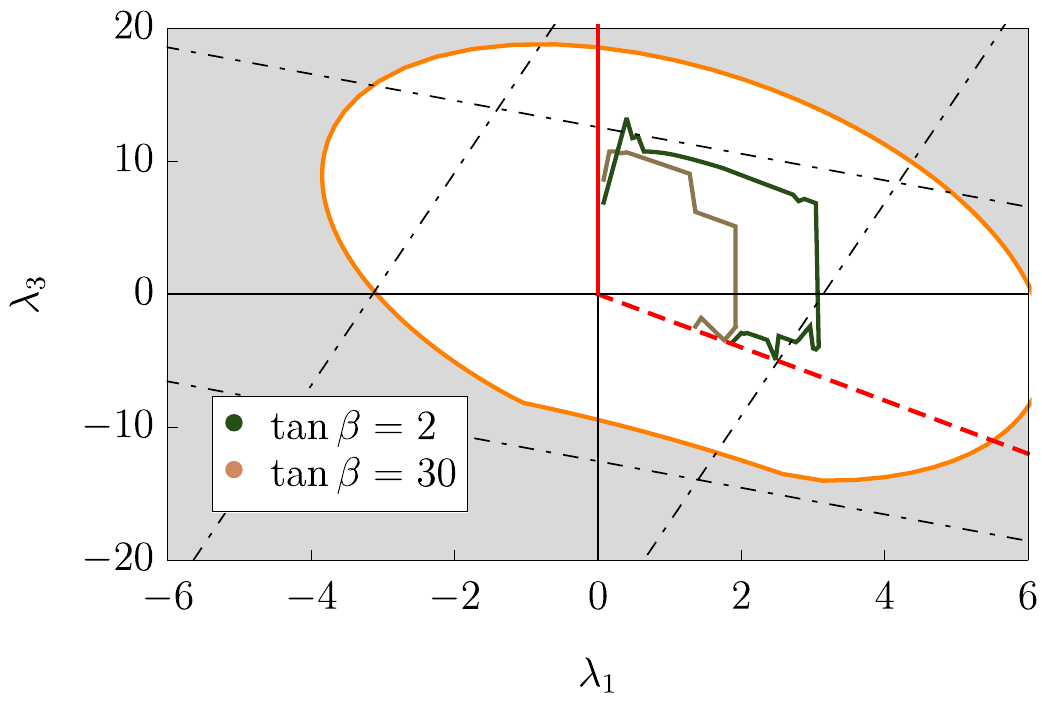}
  \caption{Tree-level and one-loop constraints on $\lambda_1$ and $\lambda_3$ in the ``MSSM-like'' THDM. Quartic-only tree-level constraints are shown as black dot-dashed lines, the vacuum stability constraint $\lambda_3 \ge - 2 \lambda_1$ is the red dashed line; our tree-level contraints including trilinears are labelled with $\tan \beta = 2$ and $\tan \beta =30$.
  The one-loop allowed region applying the constraint $|\mathrm{Re} (a_0)| \le \frac{1}{2}$ is the white region enclosed by the solid orange curve.}
\label{FIG:1La0lthalf}
\end{figure}

Hence in Fig. \ref{FIG:1La0lthalf} we recompute the constraints at one-loop applying instead $|\mathrm{Re}(a_0)| \le \frac{1}{2}$ for $\mathrm{Re}(a_0) = a_0^0 + b_{1,R}$ for the same scattering processes listed above. We use the expressions in the appendix from Ref.~\cite{Grinstein:2015rtl} to obtain the one-loop scattering amplitudes neglecting the wavefunction renormalisation contributions. These are mass-dependent and were found to be small in Ref.~\cite{Grinstein:2015rtl}. The reason is that, in the limit that $\sqrt{s}$ is much larger than all masses, only diagonal self-energies appear in the results which consist of expressions of the form
\begin{align}
z_{ii}^{1/2} \sim (v\lambda)^2 \frac{d}{ds} (B_0 (s,m^2, m^2)) \sim \frac{(v\lambda)^2}{s} \rightarrow 0. 
\end{align}
Due to the presence of the trilinear couplings in these terms they appear at the same order as box and triangle diagrams. 

The one-loop constraint is then stronger than the ``naive'' tree-level one in some cases, and weaker in others; but we find that our tree-level constraints including the effect of trilinears are stronger than both in almost all cases.

For comparison, one could also check the one-loop allowed region for the sometimes used criterion $|a_0| \le 1$. These are almost universally weaker than the tree-level constraints, implying that they are not sufficiently conservative, as can be expected.



\section{Conclusion}
\label{sec:conclusion}
We have revised the tree-level perturbative constraints in THDMs by including the contributions from (effective) trilinear couplings, and provide an extension of the package \SARAH which makes it possible to include these constraints in phenomenological studies in THDMs and many other BSM models. We found that the obtained limits can be 
significantly stronger than the ones usually applied in literature which are only correct in the limit of large scattering energies $\sqrt{s}$. The importance of the improved constraints has been demonstrated 
by two chosen examples: (i) it was shown that the values of $M_{12}$ are highly constrained in the light Higgs windows; (ii) one finds a stronger upper limit for the CP-even Higgs mass in scenarios with $\sqrt{|M_{12}|}<m_A,m_{H^+}$. On the other side, we have also discussed that the restriction to maximal scattering energies of a few TeV can revive points which violated unitarity only at much higher energies. We also made comparison with previous constraints derived at one-loop level in the large $s$ approximation. Our results indicate that the tree-level constraints including trilinear couplings are the most important for this class of models, and are not superseded by the one-loop large-momentum constraints; instead, it would be a very interesting if rather complicated task to include the effects of the trilinear couplings at one-loop order, which could potentially strengthen constraints on these models further. \\
In other BSM models similar -- or even larger -- differences between the full calculation and the large $s$ approximation can be seen. This is discussed for example in Ref.~\cite{inpreptriplet} for several triplet extensions.

\section*{Acknowledgements}
We thank Marco Sekulla for helpful discussions. FS is supported by ERC Recognition Award ERC-RA-0008 of the Helmholtz Association. MDG acknowledges support from the Agence Nationale de Recherche grant ANR-15-CE31-0002 ``HiggsAutomator'', and the Labex ``Institut Lagrange de Paris'' (ANR-11-IDEX-0004-02,  ANR-10-LABX-63). We would like to thank Sophie Williamson and Manuel Krauss for helpful discussions and collaboration on related topics.

\begin{appendix}



\end{appendix}

\bibliography{lit}

\begin{thebibliography}{40}%
\makeatletter
\providecommand \@ifxundefined [1]{%
 \@ifx{#1\undefined}
}%
\providecommand \@ifnum [1]{%
 \ifnum #1\expandafter \@firstoftwo
 \else \expandafter \@secondoftwo
 \fi
}%
\providecommand \@ifx [1]{%
 \ifx #1\expandafter \@firstoftwo
 \else \expandafter \@secondoftwo
 \fi
}%
\providecommand \natexlab [1]{#1}%
\providecommand \enquote  [1]{``#1''}%
\providecommand \bibnamefont  [1]{#1}%
\providecommand \bibfnamefont [1]{#1}%
\providecommand \citenamefont [1]{#1}%
\providecommand \href@noop [0]{\@secondoftwo}%
\providecommand \href [0]{\begingroup \@sanitize@url \@href}%
\providecommand \@href[1]{\@@startlink{#1}\@@href}%
\providecommand \@@href[1]{\endgroup#1\@@endlink}%
\providecommand \@sanitize@url [0]{\catcode `\\12\catcode `\$12\catcode
  `\&12\catcode `\#12\catcode `\^12\catcode `\_12\catcode `\%12\relax}%
\providecommand \@@startlink[1]{}%
\providecommand \@@endlink[0]{}%
\providecommand \url  [0]{\begingroup\@sanitize@url \@url }%
\providecommand \@url [1]{\endgroup\@href {#1}{\urlprefix }}%
\providecommand \urlprefix  [0]{URL }%
\providecommand \Eprint [0]{\href }%
\providecommand \doibase [0]{http://dx.doi.org/}%
\providecommand \selectlanguage [0]{\@gobble}%
\providecommand \bibinfo  [0]{\@secondoftwo}%
\providecommand \bibfield  [0]{\@secondoftwo}%
\providecommand \translation [1]{[#1]}%
\providecommand \BibitemOpen [0]{}%
\providecommand \bibitemStop [0]{}%
\providecommand \bibitemNoStop [0]{.\EOS\space}%
\providecommand \EOS [0]{\spacefactor3000\relax}%
\providecommand \BibitemShut  [1]{\csname bibitem#1\endcsname}%
\let\auto@bib@innerbib\@empty
\bibitem [{\citenamefont {Aad}\ \emph {et~al.}(2012)\citenamefont {Aad} \emph
  {et~al.}}]{Aad:2012tfa}%
  \BibitemOpen
  \bibfield  {author} {\bibinfo {author} {\bibfnamefont {G.}~\bibnamefont
  {Aad}} \emph {et~al.} (\bibinfo {collaboration} {ATLAS}),\ }\href {\doibase
  10.1016/j.physletb.2012.08.020} {\bibfield  {journal} {\bibinfo  {journal}
  {Phys. Lett.}\ }\textbf {\bibinfo {volume} {B716}},\ \bibinfo {pages} {1}
  (\bibinfo {year} {2012})},\ \Eprint {http://arxiv.org/abs/1207.7214}
  {arXiv:1207.7214 [hep-ex]} \BibitemShut {NoStop}%
\bibitem [{\citenamefont {Chatrchyan}\ \emph {et~al.}(2012)\citenamefont
  {Chatrchyan} \emph {et~al.}}]{Chatrchyan:2012xdj}%
  \BibitemOpen
  \bibfield  {author} {\bibinfo {author} {\bibfnamefont {S.}~\bibnamefont
  {Chatrchyan}} \emph {et~al.} (\bibinfo {collaboration} {CMS}),\ }\href
  {\doibase 10.1016/j.physletb.2012.08.021} {\bibfield  {journal} {\bibinfo
  {journal} {Phys. Lett.}\ }\textbf {\bibinfo {volume} {B716}},\ \bibinfo
  {pages} {30} (\bibinfo {year} {2012})},\ \Eprint
  {http://arxiv.org/abs/1207.7235} {arXiv:1207.7235 [hep-ex]} \BibitemShut
  {NoStop}%
\bibitem [{\citenamefont {Klimenko}(1985)}]{Klimenko:1984qx}%
  \BibitemOpen
  \bibfield  {author} {\bibinfo {author} {\bibfnamefont {K.~G.}\ \bibnamefont
  {Klimenko}},\ }\href {\doibase 10.1007/BF01034825} {\bibfield  {journal}
  {\bibinfo  {journal} {Theor. Math. Phys.}\ }\textbf {\bibinfo {volume}
  {62}},\ \bibinfo {pages} {58} (\bibinfo {year} {1985})},\ \bibinfo {note}
  {[Teor. Mat. Fiz.62,87(1985)]}\BibitemShut {NoStop}%
\bibitem [{\citenamefont {Velhinho}\ \emph {et~al.}(1994)\citenamefont
  {Velhinho}, \citenamefont {Santos},\ and\ \citenamefont
  {Barroso}}]{Velhinho:1994np}%
  \BibitemOpen
  \bibfield  {author} {\bibinfo {author} {\bibfnamefont {J.}~\bibnamefont
  {Velhinho}}, \bibinfo {author} {\bibfnamefont {R.}~\bibnamefont {Santos}}, \
  and\ \bibinfo {author} {\bibfnamefont {A.}~\bibnamefont {Barroso}},\ }\href
  {\doibase 10.1016/0370-2693(94)91109-6} {\bibfield  {journal} {\bibinfo
  {journal} {Phys. Lett.}\ }\textbf {\bibinfo {volume} {B322}},\ \bibinfo
  {pages} {213} (\bibinfo {year} {1994})}\BibitemShut {NoStop}%
\bibitem [{\citenamefont {Ferreira}\ \emph {et~al.}(2004)\citenamefont
  {Ferreira}, \citenamefont {Santos},\ and\ \citenamefont
  {Barroso}}]{Ferreira:2004yd}%
  \BibitemOpen
  \bibfield  {author} {\bibinfo {author} {\bibfnamefont {P.~M.}\ \bibnamefont
  {Ferreira}}, \bibinfo {author} {\bibfnamefont {R.}~\bibnamefont {Santos}}, \
  and\ \bibinfo {author} {\bibfnamefont {A.}~\bibnamefont {Barroso}},\ }\href
  {\doibase 10.1016/j.physletb.2004.10.022, 10.1016/j.physletb.2005.09.074}
  {\bibfield  {journal} {\bibinfo  {journal} {Phys. Lett.}\ }\textbf {\bibinfo
  {volume} {B603}},\ \bibinfo {pages} {219} (\bibinfo {year} {2004})},\
  \bibinfo {note} {[Erratum: Phys. Lett.B629,114(2005)]},\ \Eprint
  {http://arxiv.org/abs/hep-ph/0406231} {arXiv:hep-ph/0406231 [hep-ph]}
  \BibitemShut {NoStop}%
\bibitem [{\citenamefont {Barroso}\ \emph {et~al.}(2006)\citenamefont
  {Barroso}, \citenamefont {Ferreira},\ and\ \citenamefont
  {Santos}}]{Barroso:2005sm}%
  \BibitemOpen
  \bibfield  {author} {\bibinfo {author} {\bibfnamefont {A.}~\bibnamefont
  {Barroso}}, \bibinfo {author} {\bibfnamefont {P.~M.}\ \bibnamefont
  {Ferreira}}, \ and\ \bibinfo {author} {\bibfnamefont {R.}~\bibnamefont
  {Santos}},\ }\href {\doibase 10.1016/j.physletb.2005.11.031} {\bibfield
  {journal} {\bibinfo  {journal} {Phys. Lett.}\ }\textbf {\bibinfo {volume}
  {B632}},\ \bibinfo {pages} {684} (\bibinfo {year} {2006})},\ \Eprint
  {http://arxiv.org/abs/hep-ph/0507224} {arXiv:hep-ph/0507224 [hep-ph]}
  \BibitemShut {NoStop}%
\bibitem [{\citenamefont {Maniatis}\ \emph {et~al.}(2006)\citenamefont
  {Maniatis}, \citenamefont {von Manteuffel}, \citenamefont {Nachtmann},\ and\
  \citenamefont {Nagel}}]{Maniatis:2006fs}%
  \BibitemOpen
  \bibfield  {author} {\bibinfo {author} {\bibfnamefont {M.}~\bibnamefont
  {Maniatis}}, \bibinfo {author} {\bibfnamefont {A.}~\bibnamefont {von
  Manteuffel}}, \bibinfo {author} {\bibfnamefont {O.}~\bibnamefont
  {Nachtmann}}, \ and\ \bibinfo {author} {\bibfnamefont {F.}~\bibnamefont
  {Nagel}},\ }\href {\doibase 10.1140/epjc/s10052-006-0016-6} {\bibfield
  {journal} {\bibinfo  {journal} {Eur. Phys. J.}\ }\textbf {\bibinfo {volume}
  {C48}},\ \bibinfo {pages} {805} (\bibinfo {year} {2006})},\ \Eprint
  {http://arxiv.org/abs/hep-ph/0605184} {arXiv:hep-ph/0605184 [hep-ph]}
  \BibitemShut {NoStop}%
\bibitem [{\citenamefont {Ivanov}(2007)}]{Ivanov:2006yq}%
  \BibitemOpen
  \bibfield  {author} {\bibinfo {author} {\bibfnamefont {I.~P.}\ \bibnamefont
  {Ivanov}},\ }\href {\doibase 10.1103/PhysRevD.76.039902,
  10.1103/PhysRevD.75.035001} {\bibfield  {journal} {\bibinfo  {journal} {Phys.
  Rev.}\ }\textbf {\bibinfo {volume} {D75}},\ \bibinfo {pages} {035001}
  (\bibinfo {year} {2007})},\ \bibinfo {note} {[Erratum: Phys.
  Rev.D76,039902(2007)]},\ \Eprint {http://arxiv.org/abs/hep-ph/0609018}
  {arXiv:hep-ph/0609018 [hep-ph]} \BibitemShut {NoStop}%
\bibitem [{\citenamefont {Ivanov}(2008)}]{Ivanov:2007de}%
  \BibitemOpen
  \bibfield  {author} {\bibinfo {author} {\bibfnamefont {I.~P.}\ \bibnamefont
  {Ivanov}},\ }\href {\doibase 10.1103/PhysRevD.77.015017} {\bibfield
  {journal} {\bibinfo  {journal} {Phys. Rev.}\ }\textbf {\bibinfo {volume}
  {D77}},\ \bibinfo {pages} {015017} (\bibinfo {year} {2008})},\ \Eprint
  {http://arxiv.org/abs/0710.3490} {arXiv:0710.3490 [hep-ph]} \BibitemShut
  {NoStop}%
\bibitem [{\citenamefont {Ivanov}(2009)}]{Ivanov:2008er}%
  \BibitemOpen
  \bibfield  {author} {\bibinfo {author} {\bibfnamefont {I.~P.}\ \bibnamefont
  {Ivanov}},\ }\href@noop {} {\bibfield  {journal} {\bibinfo  {journal} {Acta
  Phys. Polon.}\ }\textbf {\bibinfo {volume} {B40}},\ \bibinfo {pages} {2789}
  (\bibinfo {year} {2009})},\ \Eprint {http://arxiv.org/abs/0812.4984}
  {arXiv:0812.4984 [hep-ph]} \BibitemShut {NoStop}%
\bibitem [{\citenamefont {Ginzburg}\ \emph {et~al.}(2010)\citenamefont
  {Ginzburg}, \citenamefont {Ivanov},\ and\ \citenamefont
  {Kanishev}}]{Ginzburg:2009dp}%
  \BibitemOpen
  \bibfield  {author} {\bibinfo {author} {\bibfnamefont {I.~F.}\ \bibnamefont
  {Ginzburg}}, \bibinfo {author} {\bibfnamefont {I.~P.}\ \bibnamefont
  {Ivanov}}, \ and\ \bibinfo {author} {\bibfnamefont {K.~A.}\ \bibnamefont
  {Kanishev}},\ }\href {\doibase 10.1103/PhysRevD.81.085031} {\bibfield
  {journal} {\bibinfo  {journal} {Phys. Rev.}\ }\textbf {\bibinfo {volume}
  {D81}},\ \bibinfo {pages} {085031} (\bibinfo {year} {2010})},\ \Eprint
  {http://arxiv.org/abs/0911.2383} {arXiv:0911.2383 [hep-ph]} \BibitemShut
  {NoStop}%
\bibitem [{\citenamefont {Staub}(2017)}]{Staub:2017ktc}%
  \BibitemOpen
  \bibfield  {author} {\bibinfo {author} {\bibfnamefont {F.}~\bibnamefont
  {Staub}},\ }\href@noop {} {\  (\bibinfo {year} {2017})},\ \Eprint
  {http://arxiv.org/abs/1705.03677} {arXiv:1705.03677 [hep-ph]} \BibitemShut
  {NoStop}%
\bibitem [{\citenamefont {Casalbuoni}\ \emph {et~al.}(1986)\citenamefont
  {Casalbuoni}, \citenamefont {Dominici}, \citenamefont {Gatto},\ and\
  \citenamefont {Giunti}}]{Casalbuoni:1986hy}%
  \BibitemOpen
  \bibfield  {author} {\bibinfo {author} {\bibfnamefont {R.}~\bibnamefont
  {Casalbuoni}}, \bibinfo {author} {\bibfnamefont {D.}~\bibnamefont
  {Dominici}}, \bibinfo {author} {\bibfnamefont {R.}~\bibnamefont {Gatto}}, \
  and\ \bibinfo {author} {\bibfnamefont {C.}~\bibnamefont {Giunti}},\
  }\bibfield  {booktitle} {\emph {\bibinfo {booktitle} {{23RD International
  Conference on High Energy Physics}}},\ }\href {\doibase
  10.1016/0370-2693(86)91502-9} {\bibfield  {journal} {\bibinfo  {journal}
  {Phys. Lett.}\ }\textbf {\bibinfo {volume} {B178}},\ \bibinfo {pages} {235}
  (\bibinfo {year} {1986})}\BibitemShut {NoStop}%
\bibitem [{\citenamefont {Casalbuoni}\ \emph {et~al.}(1988)\citenamefont
  {Casalbuoni}, \citenamefont {Dominici}, \citenamefont {Feruglio},\ and\
  \citenamefont {Gatto}}]{Casalbuoni:1987eg}%
  \BibitemOpen
  \bibfield  {author} {\bibinfo {author} {\bibfnamefont {R.}~\bibnamefont
  {Casalbuoni}}, \bibinfo {author} {\bibfnamefont {D.}~\bibnamefont
  {Dominici}}, \bibinfo {author} {\bibfnamefont {F.}~\bibnamefont {Feruglio}},
  \ and\ \bibinfo {author} {\bibfnamefont {R.}~\bibnamefont {Gatto}},\ }\href
  {\doibase 10.1016/0370-2693(88)90158-X} {\bibfield  {journal} {\bibinfo
  {journal} {Phys. Lett.}\ }\textbf {\bibinfo {volume} {B200}},\ \bibinfo
  {pages} {495} (\bibinfo {year} {1988})}\BibitemShut {NoStop}%
\bibitem [{\citenamefont {Maalampi}\ \emph {et~al.}(1991)\citenamefont
  {Maalampi}, \citenamefont {Sirkka},\ and\ \citenamefont
  {Vilja}}]{Maalampi:1991fb}%
  \BibitemOpen
  \bibfield  {author} {\bibinfo {author} {\bibfnamefont {J.}~\bibnamefont
  {Maalampi}}, \bibinfo {author} {\bibfnamefont {J.}~\bibnamefont {Sirkka}}, \
  and\ \bibinfo {author} {\bibfnamefont {I.}~\bibnamefont {Vilja}},\ }\href
  {\doibase 10.1016/0370-2693(91)90068-2} {\bibfield  {journal} {\bibinfo
  {journal} {Phys. Lett.}\ }\textbf {\bibinfo {volume} {B265}},\ \bibinfo
  {pages} {371} (\bibinfo {year} {1991})}\BibitemShut {NoStop}%
\bibitem [{\citenamefont {Kanemura}\ \emph {et~al.}(1993)\citenamefont
  {Kanemura}, \citenamefont {Kubota},\ and\ \citenamefont
  {Takasugi}}]{Kanemura:1993hm}%
  \BibitemOpen
  \bibfield  {author} {\bibinfo {author} {\bibfnamefont {S.}~\bibnamefont
  {Kanemura}}, \bibinfo {author} {\bibfnamefont {T.}~\bibnamefont {Kubota}}, \
  and\ \bibinfo {author} {\bibfnamefont {E.}~\bibnamefont {Takasugi}},\ }\href
  {\doibase 10.1016/0370-2693(93)91205-2} {\bibfield  {journal} {\bibinfo
  {journal} {Phys. Lett.}\ }\textbf {\bibinfo {volume} {B313}},\ \bibinfo
  {pages} {155} (\bibinfo {year} {1993})},\ \Eprint
  {http://arxiv.org/abs/hep-ph/9303263} {arXiv:hep-ph/9303263 [hep-ph]}
  \BibitemShut {NoStop}%
\bibitem [{\citenamefont {Ginzburg}\ and\ \citenamefont
  {Ivanov}(2003)}]{Ginzburg:2003fe}%
  \BibitemOpen
  \bibfield  {author} {\bibinfo {author} {\bibfnamefont {I.~F.}\ \bibnamefont
  {Ginzburg}}\ and\ \bibinfo {author} {\bibfnamefont {I.~P.}\ \bibnamefont
  {Ivanov}},\ }\href@noop {} {\  (\bibinfo {year} {2003})},\ \Eprint
  {http://arxiv.org/abs/hep-ph/0312374} {arXiv:hep-ph/0312374 [hep-ph]}
  \BibitemShut {NoStop}%
\bibitem [{\citenamefont {Akeroyd}\ \emph {et~al.}(2000)\citenamefont
  {Akeroyd}, \citenamefont {Arhrib},\ and\ \citenamefont
  {Naimi}}]{Akeroyd:2000wc}%
  \BibitemOpen
  \bibfield  {author} {\bibinfo {author} {\bibfnamefont {A.~G.}\ \bibnamefont
  {Akeroyd}}, \bibinfo {author} {\bibfnamefont {A.}~\bibnamefont {Arhrib}}, \
  and\ \bibinfo {author} {\bibfnamefont {E.-M.}\ \bibnamefont {Naimi}},\ }\href
  {\doibase 10.1016/S0370-2693(00)00962-X} {\bibfield  {journal} {\bibinfo
  {journal} {Phys. Lett.}\ }\textbf {\bibinfo {volume} {B490}},\ \bibinfo
  {pages} {119} (\bibinfo {year} {2000})},\ \Eprint
  {http://arxiv.org/abs/hep-ph/0006035} {arXiv:hep-ph/0006035 [hep-ph]}
  \BibitemShut {NoStop}%
\bibitem [{\citenamefont {Horejsi}\ and\ \citenamefont
  {Kladiva}(2006)}]{Horejsi:2005da}%
  \BibitemOpen
  \bibfield  {author} {\bibinfo {author} {\bibfnamefont {J.}~\bibnamefont
  {Horejsi}}\ and\ \bibinfo {author} {\bibfnamefont {M.}~\bibnamefont
  {Kladiva}},\ }\href {\doibase 10.1140/epjc/s2006-02472-3} {\bibfield
  {journal} {\bibinfo  {journal} {Eur. Phys. J.}\ }\textbf {\bibinfo {volume}
  {C46}},\ \bibinfo {pages} {81} (\bibinfo {year} {2006})},\ \Eprint
  {http://arxiv.org/abs/hep-ph/0510154} {arXiv:hep-ph/0510154 [hep-ph]}
  \BibitemShut {NoStop}%
\bibitem [{\citenamefont {Cynolter}\ \emph {et~al.}(2005)\citenamefont
  {Cynolter}, \citenamefont {Lendvai},\ and\ \citenamefont
  {Pocsik}}]{Cynolter:2004cq}%
  \BibitemOpen
  \bibfield  {author} {\bibinfo {author} {\bibfnamefont {G.}~\bibnamefont
  {Cynolter}}, \bibinfo {author} {\bibfnamefont {E.}~\bibnamefont {Lendvai}}, \
  and\ \bibinfo {author} {\bibfnamefont {G.}~\bibnamefont {Pocsik}},\
  }\href@noop {} {\bibfield  {journal} {\bibinfo  {journal} {Acta Phys.
  Polon.}\ }\textbf {\bibinfo {volume} {B36}},\ \bibinfo {pages} {827}
  (\bibinfo {year} {2005})},\ \Eprint {http://arxiv.org/abs/hep-ph/0410102}
  {arXiv:hep-ph/0410102 [hep-ph]} \BibitemShut {NoStop}%
\bibitem [{\citenamefont {Kang}\ and\ \citenamefont
  {Park}(2015)}]{Kang:2013zba}%
  \BibitemOpen
  \bibfield  {author} {\bibinfo {author} {\bibfnamefont {S.~K.}\ \bibnamefont
  {Kang}}\ and\ \bibinfo {author} {\bibfnamefont {J.}~\bibnamefont {Park}},\
  }\href {\doibase 10.1007/JHEP04(2015)009} {\bibfield  {journal} {\bibinfo
  {journal} {JHEP}\ }\textbf {\bibinfo {volume} {04}},\ \bibinfo {pages} {009}
  (\bibinfo {year} {2015})},\ \Eprint {http://arxiv.org/abs/1306.6713}
  {arXiv:1306.6713 [hep-ph]} \BibitemShut {NoStop}%
\bibitem [{\citenamefont {Schuessler}\ and\ \citenamefont
  {Zeppenfeld}(2007)}]{Schuessler:2007av}%
  \BibitemOpen
  \bibfield  {author} {\bibinfo {author} {\bibfnamefont {A.}~\bibnamefont
  {Schuessler}}\ and\ \bibinfo {author} {\bibfnamefont {D.}~\bibnamefont
  {Zeppenfeld}},\ }in\ \href
  {http://www.susy07.uni-karlsruhe.de/Proceedings/proceedings/susy07.pdf}
  {\emph {\bibinfo {booktitle} {{SUSY 2007 Proceedings}}}}\ (\bibinfo {year}
  {2007})\ pp.\ \bibinfo {pages} {236--239},\ \Eprint
  {http://arxiv.org/abs/0710.5175} {arXiv:0710.5175 [hep-ph]} \BibitemShut
  {NoStop}%
\bibitem [{\citenamefont {Porod}(2003)}]{Porod:2003um}%
  \BibitemOpen
  \bibfield  {author} {\bibinfo {author} {\bibfnamefont {W.}~\bibnamefont
  {Porod}},\ }\href {\doibase 10.1016/S0010-4655(03)00222-4} {\bibfield
  {journal} {\bibinfo  {journal} {Comput.Phys.Commun.}\ }\textbf {\bibinfo
  {volume} {153}},\ \bibinfo {pages} {275} (\bibinfo {year} {2003})},\ \Eprint
  {http://arxiv.org/abs/hep-ph/0301101} {arXiv:hep-ph/0301101 [hep-ph]}
  \BibitemShut {NoStop}%
\bibitem [{\citenamefont {Porod}\ and\ \citenamefont
  {Staub}(2011)}]{Porod:2011nf}%
  \BibitemOpen
  \bibfield  {author} {\bibinfo {author} {\bibfnamefont {W.}~\bibnamefont
  {Porod}}\ and\ \bibinfo {author} {\bibfnamefont {F.}~\bibnamefont {Staub}},\
  }\href@noop {} {\  (\bibinfo {year} {2011})},\ \Eprint
  {http://arxiv.org/abs/1104.1573} {arXiv:1104.1573 [hep-ph]} \BibitemShut
  {NoStop}%
\bibitem [{\citenamefont {Goodsell}\ and\ \citenamefont
  {Staub}(rxiv)}]{inprepsarah}%
  \BibitemOpen
  \bibfield  {author} {\bibinfo {author} {\bibfnamefont {M.}~\bibnamefont
  {Goodsell}}\ and\ \bibinfo {author} {\bibfnamefont {F.}~\bibnamefont
  {Staub}},\ }\href@noop {} {\  (\bibinfo {year} {on arxiv})}\BibitemShut
  {NoStop}%
\bibitem [{\citenamefont {Misiak}\ and\ \citenamefont
  {Steinhauser}(2017)}]{Misiak:2017bgg}%
  \BibitemOpen
  \bibfield  {author} {\bibinfo {author} {\bibfnamefont {M.}~\bibnamefont
  {Misiak}}\ and\ \bibinfo {author} {\bibfnamefont {M.}~\bibnamefont
  {Steinhauser}},\ }\href {\doibase 10.1140/epjc/s10052-017-4776-y} {\bibfield
  {journal} {\bibinfo  {journal} {Eur. Phys. J.}\ }\textbf {\bibinfo {volume}
  {C77}},\ \bibinfo {pages} {201} (\bibinfo {year} {2017})},\ \Eprint
  {http://arxiv.org/abs/1702.04571} {arXiv:1702.04571 [hep-ph]} \BibitemShut
  {NoStop}%
\bibitem [{\citenamefont {Bechtle}\ \emph {et~al.}(2010)\citenamefont
  {Bechtle}, \citenamefont {Brein}, \citenamefont {Heinemeyer}, \citenamefont
  {Weiglein},\ and\ \citenamefont {Williams}}]{Bechtle:2008jh}%
  \BibitemOpen
  \bibfield  {author} {\bibinfo {author} {\bibfnamefont {P.}~\bibnamefont
  {Bechtle}}, \bibinfo {author} {\bibfnamefont {O.}~\bibnamefont {Brein}},
  \bibinfo {author} {\bibfnamefont {S.}~\bibnamefont {Heinemeyer}}, \bibinfo
  {author} {\bibfnamefont {G.}~\bibnamefont {Weiglein}}, \ and\ \bibinfo
  {author} {\bibfnamefont {K.~E.}\ \bibnamefont {Williams}},\ }\href {\doibase
  10.1016/j.cpc.2009.09.003} {\bibfield  {journal} {\bibinfo  {journal}
  {Comput. Phys. Commun.}\ }\textbf {\bibinfo {volume} {181}},\ \bibinfo
  {pages} {138} (\bibinfo {year} {2010})},\ \Eprint
  {http://arxiv.org/abs/0811.4169} {arXiv:0811.4169 [hep-ph]} \BibitemShut
  {NoStop}%
\bibitem [{\citenamefont {Bechtle}\ \emph {et~al.}(2011)\citenamefont
  {Bechtle}, \citenamefont {Brein}, \citenamefont {Heinemeyer}, \citenamefont
  {Weiglein},\ and\ \citenamefont {Williams}}]{Bechtle:2011sb}%
  \BibitemOpen
  \bibfield  {author} {\bibinfo {author} {\bibfnamefont {P.}~\bibnamefont
  {Bechtle}}, \bibinfo {author} {\bibfnamefont {O.}~\bibnamefont {Brein}},
  \bibinfo {author} {\bibfnamefont {S.}~\bibnamefont {Heinemeyer}}, \bibinfo
  {author} {\bibfnamefont {G.}~\bibnamefont {Weiglein}}, \ and\ \bibinfo
  {author} {\bibfnamefont {K.~E.}\ \bibnamefont {Williams}},\ }\href {\doibase
  10.1016/j.cpc.2011.07.015} {\bibfield  {journal} {\bibinfo  {journal}
  {Comput.Phys.Commun.}\ }\textbf {\bibinfo {volume} {182}},\ \bibinfo {pages}
  {2605} (\bibinfo {year} {2011})},\ \Eprint {http://arxiv.org/abs/1102.1898}
  {arXiv:1102.1898 [hep-ph]} \BibitemShut {NoStop}%
\bibitem [{\citenamefont {Bechtle}\ \emph {et~al.}(2014)\citenamefont
  {Bechtle}, \citenamefont {Brein}, \citenamefont {Heinemeyer}, \citenamefont
  {Stal}, \citenamefont {Stefaniak}, \citenamefont {Weiglein},\ and\
  \citenamefont {Williams}}]{Bechtle:2013wla}%
  \BibitemOpen
  \bibfield  {author} {\bibinfo {author} {\bibfnamefont {P.}~\bibnamefont
  {Bechtle}}, \bibinfo {author} {\bibfnamefont {O.}~\bibnamefont {Brein}},
  \bibinfo {author} {\bibfnamefont {S.}~\bibnamefont {Heinemeyer}}, \bibinfo
  {author} {\bibfnamefont {O.}~\bibnamefont {Stal}}, \bibinfo {author}
  {\bibfnamefont {T.}~\bibnamefont {Stefaniak}}, \bibinfo {author}
  {\bibfnamefont {G.}~\bibnamefont {Weiglein}}, \ and\ \bibinfo {author}
  {\bibfnamefont {K.~E.}\ \bibnamefont {Williams}},\ }\href {\doibase
  10.1140/epjc/s10052-013-2693-2} {\bibfield  {journal} {\bibinfo  {journal}
  {Eur. Phys. J.}\ }\textbf {\bibinfo {volume} {C74}},\ \bibinfo {pages} {2693}
  (\bibinfo {year} {2014})},\ \Eprint {http://arxiv.org/abs/1311.0055}
  {arXiv:1311.0055 [hep-ph]} \BibitemShut {NoStop}%
\bibitem [{\citenamefont {Staub}(2008)}]{Staub:2008uz}%
  \BibitemOpen
  \bibfield  {author} {\bibinfo {author} {\bibfnamefont {F.}~\bibnamefont
  {Staub}},\ }\href@noop {} {\  (\bibinfo {year} {2008})},\ \Eprint
  {http://arxiv.org/abs/0806.0538} {arXiv:0806.0538 [hep-ph]} \BibitemShut
  {NoStop}%
\bibitem [{\citenamefont {Staub}(2010)}]{Staub:2009bi}%
  \BibitemOpen
  \bibfield  {author} {\bibinfo {author} {\bibfnamefont {F.}~\bibnamefont
  {Staub}},\ }\href {\doibase 10.1016/j.cpc.2010.01.011} {\bibfield  {journal}
  {\bibinfo  {journal} {Comput.Phys.Commun.}\ }\textbf {\bibinfo {volume}
  {181}},\ \bibinfo {pages} {1077} (\bibinfo {year} {2010})},\ \Eprint
  {http://arxiv.org/abs/0909.2863} {arXiv:0909.2863 [hep-ph]} \BibitemShut
  {NoStop}%
\bibitem [{\citenamefont {Staub}(2011)}]{Staub:2010jh}%
  \BibitemOpen
  \bibfield  {author} {\bibinfo {author} {\bibfnamefont {F.}~\bibnamefont
  {Staub}},\ }\href {\doibase 10.1016/j.cpc.2010.11.030} {\bibfield  {journal}
  {\bibinfo  {journal} {Comput.Phys.Commun.}\ }\textbf {\bibinfo {volume}
  {182}},\ \bibinfo {pages} {808} (\bibinfo {year} {2011})},\ \Eprint
  {http://arxiv.org/abs/1002.0840} {arXiv:1002.0840 [hep-ph]} \BibitemShut
  {NoStop}%
\bibitem [{\citenamefont {Staub}(2012)}]{Staub:2012pb}%
  \BibitemOpen
  \bibfield  {author} {\bibinfo {author} {\bibfnamefont {F.}~\bibnamefont
  {Staub}},\ }\href@noop {} {\  (\bibinfo {year} {2012})},\ \Eprint
  {http://arxiv.org/abs/1207.0906} {arXiv:1207.0906 [hep-ph]} \BibitemShut
  {NoStop}%
\bibitem [{\citenamefont {Staub}(2014)}]{Staub:2013tta}%
  \BibitemOpen
  \bibfield  {author} {\bibinfo {author} {\bibfnamefont {F.}~\bibnamefont
  {Staub}},\ }\href {\doibase 10.1016/j.cpc.2014.02.018} {\bibfield  {journal}
  {\bibinfo  {journal} {Comput. Phys. Commun.}\ }\textbf {\bibinfo {volume}
  {185}},\ \bibinfo {pages} {1773} (\bibinfo {year} {2014})},\ \Eprint
  {http://arxiv.org/abs/1309.7223} {arXiv:1309.7223 [hep-ph]} \BibitemShut
  {NoStop}%
\bibitem [{\citenamefont {Goodsell}\ \emph
  {et~al.}(2015{\natexlab{a}})\citenamefont {Goodsell}, \citenamefont
  {Nickel},\ and\ \citenamefont {Staub}}]{Goodsell:2014bna}%
  \BibitemOpen
  \bibfield  {author} {\bibinfo {author} {\bibfnamefont {M.~D.}\ \bibnamefont
  {Goodsell}}, \bibinfo {author} {\bibfnamefont {K.}~\bibnamefont {Nickel}}, \
  and\ \bibinfo {author} {\bibfnamefont {F.}~\bibnamefont {Staub}},\ }\href
  {\doibase 10.1140/epjc/s10052-014-3247-y} {\bibfield  {journal} {\bibinfo
  {journal} {Eur. Phys. J.}\ }\textbf {\bibinfo {volume} {C75}},\ \bibinfo
  {pages} {32} (\bibinfo {year} {2015}{\natexlab{a}})},\ \Eprint
  {http://arxiv.org/abs/1411.0675} {arXiv:1411.0675 [hep-ph]} \BibitemShut
  {NoStop}%
\bibitem [{\citenamefont {Goodsell}\ \emph
  {et~al.}(2015{\natexlab{b}})\citenamefont {Goodsell}, \citenamefont
  {Nickel},\ and\ \citenamefont {Staub}}]{Goodsell:2015ira}%
  \BibitemOpen
  \bibfield  {author} {\bibinfo {author} {\bibfnamefont {M.}~\bibnamefont
  {Goodsell}}, \bibinfo {author} {\bibfnamefont {K.}~\bibnamefont {Nickel}}, \
  and\ \bibinfo {author} {\bibfnamefont {F.}~\bibnamefont {Staub}},\ }\href
  {\doibase 10.1140/epjc/s10052-015-3494-6} {\bibfield  {journal} {\bibinfo
  {journal} {Eur. Phys. J.}\ }\textbf {\bibinfo {volume} {C75}},\ \bibinfo
  {pages} {290} (\bibinfo {year} {2015}{\natexlab{b}})},\ \Eprint
  {http://arxiv.org/abs/1503.03098} {arXiv:1503.03098 [hep-ph]} \BibitemShut
  {NoStop}%
\bibitem [{\citenamefont {Braathen}\ \emph {et~al.}(2017)\citenamefont
  {Braathen}, \citenamefont {Goodsell},\ and\ \citenamefont
  {Staub}}]{Braathen:2017izn}%
  \BibitemOpen
  \bibfield  {author} {\bibinfo {author} {\bibfnamefont {J.}~\bibnamefont
  {Braathen}}, \bibinfo {author} {\bibfnamefont {M.~D.}\ \bibnamefont
  {Goodsell}}, \ and\ \bibinfo {author} {\bibfnamefont {F.}~\bibnamefont
  {Staub}},\ }\href {\doibase 10.1140/epjc/s10052-017-5303-x} {\bibfield
  {journal} {\bibinfo  {journal} {Eur. Phys. J.}\ }\textbf {\bibinfo {volume}
  {C77}},\ \bibinfo {pages} {757} (\bibinfo {year} {2017})},\ \Eprint
  {http://arxiv.org/abs/1706.05372} {arXiv:1706.05372 [hep-ph]} \BibitemShut
  {NoStop}%
\bibitem [{\citenamefont {Krauss}\ and\ \citenamefont
  {Staub}(2018)}]{Krauss:2017xpj}%
  \BibitemOpen
  \bibfield  {author} {\bibinfo {author} {\bibfnamefont {M.~E.}\ \bibnamefont
  {Krauss}}\ and\ \bibinfo {author} {\bibfnamefont {F.}~\bibnamefont {Staub}},\
  }\href {\doibase 10.1140/epjc/s10052-018-5676-5} {\bibfield  {journal}
  {\bibinfo  {journal} {Eur. Phys. J.}\ }\textbf {\bibinfo {volume} {C78}},\
  \bibinfo {pages} {185} (\bibinfo {year} {2018})},\ \Eprint
  {http://arxiv.org/abs/1709.03501} {arXiv:1709.03501 [hep-ph]} \BibitemShut
  {NoStop}%
\bibitem [{\citenamefont {Grinstein}\ \emph {et~al.}(2016)\citenamefont
  {Grinstein}, \citenamefont {Murphy},\ and\ \citenamefont
  {Uttayarat}}]{Grinstein:2015rtl}%
  \BibitemOpen
  \bibfield  {author} {\bibinfo {author} {\bibfnamefont {B.}~\bibnamefont
  {Grinstein}}, \bibinfo {author} {\bibfnamefont {C.~W.}\ \bibnamefont
  {Murphy}}, \ and\ \bibinfo {author} {\bibfnamefont {P.}~\bibnamefont
  {Uttayarat}},\ }\href {\doibase 10.1007/JHEP06(2016)070} {\bibfield
  {journal} {\bibinfo  {journal} {JHEP}\ }\textbf {\bibinfo {volume} {06}},\
  \bibinfo {pages} {070} (\bibinfo {year} {2016})},\ \Eprint
  {http://arxiv.org/abs/1512.04567} {arXiv:1512.04567 [hep-ph]} \BibitemShut
  {NoStop}%
\bibitem [{\citenamefont {Krauss}\ and\ \citenamefont
  {Staub}(rxiv)}]{inpreptriplet}%
  \BibitemOpen
  \bibfield  {author} {\bibinfo {author} {\bibfnamefont {M.}~\bibnamefont
  {Krauss}}\ and\ \bibinfo {author} {\bibfnamefont {F.}~\bibnamefont {Staub}},\
  }\href@noop {} {\  (\bibinfo {year} {on arxiv})}\BibitemShut {NoStop}%
\end{thebibliography}%

\end{document}